\documentclass[
preprint,
 amsmath,amssymb,
 aip, jcp
]{revtex4-1}

\usepackage{graphicx}
\usepackage{dcolumn}
\usepackage{bm}
\usepackage{xcolor}

\usepackage{amsmath}
\usepackage{amssymb}

\begin{document}

\title{Solvation free-energy pressure corrections in the Three Dimensional Reference Interaction Site Model}

\author{Volodymyr Sergiievskyi}
	\email{voov.rat@gmail.com}
	\affiliation{University Duisburg-Essen, Universit\"atsstra{\ss}e 5
45117 Essen, Germany}

\author{Guillaume Jeanmairet}
	\affiliation{Max Planck Institute F\"ur Kernphysik,
Electronic Structure Theory
Heisenbergstra{\ss}e 1,
Stuttgart 70569,
Germany }

\author{Maximilien Levesque}
	\affiliation{\'Ecole Normale Sup\'erieure, PSL Research University, D\'epartement de Chimie, Sorbonne Universit\'es -- UPMC Univ Paris 06, CNRS UMR 8640 PASTEUR, 24 rue Lhomond, 75005 Paris, France}

\author{Daniel Borgis}
	\affiliation{\'Ecole Normale Sup\'erieure, PSL Research University, D\'epartement de Chimie, Sorbonne Universit\'es -- UPMC Univ Paris 06, CNRS UMR 8640 PASTEUR, 24 rue Lhomond, 75005 Paris, France}
	\affiliation{Maison de la Simulation, USR 3441, CEA-CNRS-INRIA- Univ. Paris-Sud - Univ. de
Versailles, 91191, Gif-sur-Yvette, France}

\begin{abstract}

Solvation free energies are efficiently predicted by molecular density functionnal theory (MDFT) if one corrects the overpressure introduced by the usual homogeneous reference fluid approximation. Sergiievskyi et al. [Sergiievskyi et al., JPCL, 2014, 5, 1935-1942] recently derived the rigorous compensation of this excess of pressure (PC) and proposed an empirical "ideal gas" supplementary correction (PC+) that further enhances the calculated solvation free energies.
In a recent paper [Misin et al,  JCP, 2015, 142, 091105], those corrections were applied to solvation free energy calculations using the three-dimensional reference interaction site model (3D-RISM). As for classical DFT, PC and PC+ corrections improve greatly the predictions of 3D-RISM, but PC+ is described as decreasing the accuracy.
In this article, we first derive rigorously the PC and PC+ corrections for 3D-RISM. We show the reported discrepancy is then taken off by introducing the correct expression of the pressure in 3D-RISM. 
This provides a consistent way to correct the solvation free-energies calculated by 3D-RISM method.

\end{abstract}

\maketitle

\section{Introduction}

 The knowledge of the solvation free energy (SFE) allows one to predict the behavior of substances in solution.
Classical density functional theory (DFT) is a perspective method for SFE calculations which in one hand can reproduce reasonably well the microscopic structural properties of the solvent and in the other hand is two to four orders of magnitude faster than all-atoms simulations.
SFE in classical DFT is calculated by minimizing a free-energy functional of the solvent density distribution only \cite{evans_nature_1979,hansen_theory_2000}.
This requires only moderate computational effort. For many systems of interest the calculations takes less than a minute on a standard computer\cite{gendre_classical_2009,sergiievskyi_3drism_2012}. 
However, despite their attractiveness, the DFT and related integral equation (IE) methods were not used for the SFE calculations until recently because impaired by computational errors.
For example, it was reported that one of the most popular hyper-netted chain (HNC) approximation dramatically overestimates the SFE, sometimes by 200-300 \

Recently it was shown also that the classical DFT and related 3D-RISM methods can be corrected by using  empirical partial molar volume (PMV) corrections
\cite{palmer_accurate_2010, palmer_toward_2011}. However, the question of their universality and transferability is still open. 
 In a recent paper, we gave a physically-based rationale for the PMV corrections within the classical DFT formalism\cite{sergiievskyi_fast_2014}.
In that paper we considered two variants of the correction, namely the pure pressure correction (PC) that rigorously compensate the overpressure due to the HNC (or HRF) approximations and the modified pressure correction (PC+) which contains an additional (and at this stage ad-hoc) term.
Numerical results for a wide range (500+) of solutes suggest that the PC+ correction is more accurate, and  that it could be used in practical applications.
The same conclusions are supported by a series of  independent investigators who used the PC+ correction in their calculations \cite{li_uranyl_2015}.

In a recent paper, Misin et al. tested the applicability  of both the PC and PC+ corrections  to  3D-RISM calculations for a large set of organic solutes of various nature\cite{misin_communication:_2015}.
They come to the paradoxical conclusion that in the 3D-RISM case the PC correction leads to more accurate results than the PC+. In that paper, however, no satisfactory explanations of this paradox are given.

In the present paper  we show that the difference in the 3D-RISM and classical DFT results can be explained  by the difference in expressions of the pressure in these models. This hypothesis is further checked by performing a series of SFE calculations for a data set of molecules and for a model hydrophobic solute.
In the next section, we derive the formal definition of the pressure correction (PC) and modified pressure correction (PC+) in the MDFT framework from new arguments. Then, we discuss SFE calculations using both MDFT and 3D-RISM.
Finally, we derive the 3D-RISM equations in the functional form and give the expression for the pressure in the 3D-RISM model. This allows us to derive the correct PC and PC+ expressions for the 3D-RISM method and to explain the discrepancy previously reported between MDFT and 3DRISM results. 

\section{Pressure correction in Classical DFT}

\label{sec:PC_in_DFT}

\begin{figure}
\includegraphics[width=1\columnwidth]{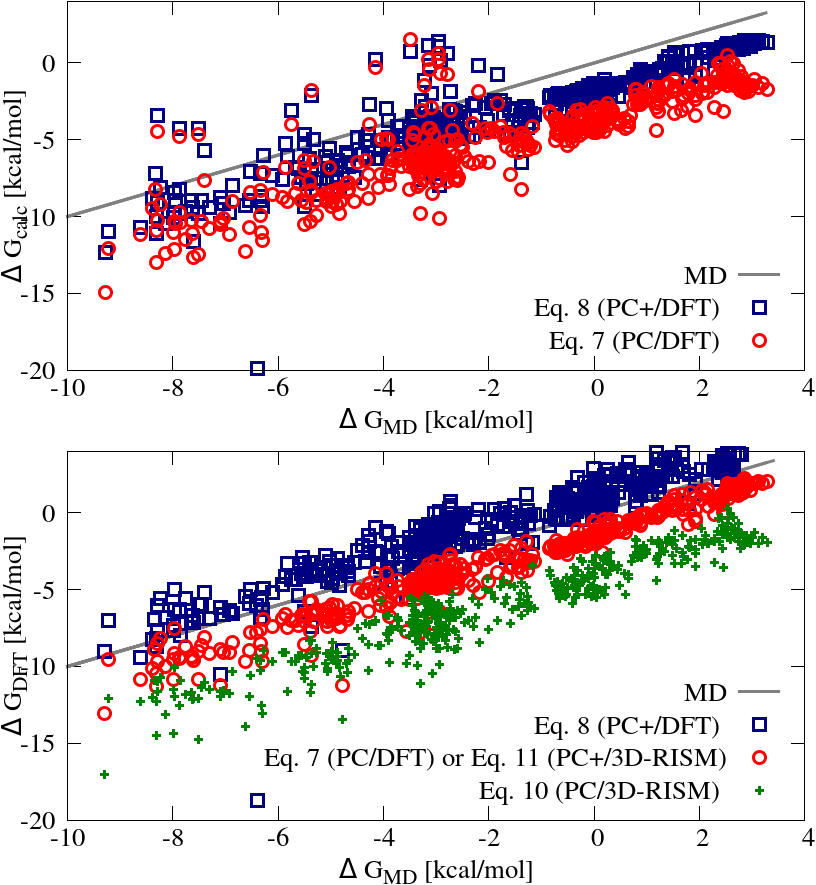}
\caption{
\label{Classical DFT_vs_3D-RISM}
Comparison of the solvation free energies calculated with Classical DFT (top) and 3D-RISM (bottom) with previously derived formulae PC (red circles) and PC+ (blue squares) corrections for a large set of small organic molecules with respect to the reference molecular dynamics (MD) results. The same correction gives systematically different results.
}
\end{figure}

\begin{table}
\begin{center}
\begin{tabular}{l | c c | c c }
Solvation free energy formula&
$\mu_{\rm DFT}$ &
$\sigma_{\rm DFT}$ &
$\mu_{\rm RISM}$ &
$\sigma_{\rm RISM}$ \\
\hline
$\Delta {\rm G}_{\rm HNC}$ &    12.952&     5.362&    15.976&     4.614\\
$\Delta {\rm G}_{\rm HNC}+\Delta {\rm V}({\rm P}_{\rm DFT}-{\rm P}_{\rm ID})$ &    -1.469&     1.376&     1.175&     1.170\\
$\Delta {\rm G}_{\rm HNC}+\Delta {\rm V}\cdot{\rm P}_{\rm DFT}$ &    -3.238&     1.651&    -1.500&     0.972\\
$\Delta {\rm G}_{\rm HNC}+\Delta {\rm V}({\rm P}_{\rm DFT}+{\rm P}_{\rm ID})$ &         -&         -&    -4.175&     1.222\\

\hline
\end{tabular}
\end{center}
\caption{
\label{tab:Classical DFT_vs_3DRISM}
Mean error in kcal/mol, $\mu$, and standard deviation in (kcal/mol)$^{2}$, $\sigma$, for the solvation free energies computed with Classical DFT and 3DRISM methods and corrected using different formulae. 
Here $P_{\rm DFT}$ is defined as in Eq. \eqref{eq:PHNC}, $P_{\rm ID} \equiv \rho_0 kT$.
}
\end{table}

We consider the process of solvating a rigid solute in the
isothermal-isobaric ($NPT$) ensemble. 
We define the volume change of the liquid system as the solute partial molar volume $\Delta V$. 
The Gibbs solvation free energy $\Delta G$ can be written as 
\begin{equation}
\label{eq:DeltaG}
   \Delta G = \Delta U - T\Delta S  + P\Delta V  .
\end{equation}
Although derived in Ref.\cite{sergiievskyi_fast_2014} through a number of equations in different thermodynamic ensemble, the pressure correction  in the calculation of $\Delta G$ by classical DFT can be rephrased with very simple arguments. If we assume that classical DFT can reproduce reasonably well the solvent structure around the solute,  we can expect that the structure-dependent components of the solvation free energy, $\Delta U$ and $T \Delta S$,  are calculated with reasonable accuracy.
On the other hand, it is known that the models based on the homogeneous reference fluid approximation (HRF, or equivalently HNC approximation in integral equations) fail to predict the pressure. Consequently, the $P\Delta V$ term is incorrect  \cite{chuev_improved_2007,sergiievskyi_fast_2014}.
The pressure correction, PC,  boils down to  eliminate the inaccurate $P\Delta V$ term of DFT and to replace it with the correct, experimental pressure term. Accounting furthermore for the fact that density functional theory is formulated in the grand-canonical ensemble with the grand potential $\Omega$, and that the solvation process implies
$\Delta G = \Delta \Omega$, we simply replace the wrong pressure by the experimental pressure.
\begin{eqnarray} 
\label{eq:PC}
  \Delta G &\approx &\Delta \Omega_{\rm DFT} - P_{\rm DFT} \Delta V_{\rm DFT} 
   + P_{\rm exp} \Delta V_{\rm exp}  .
\end{eqnarray}
In  experiments or liquid simulations performed at atmospheric pressure (1 atm $\approx 10^{-5}$ kcal/mol per \AA$^3$), the $P_{\rm exp} \Delta V_{\rm exp}$ term is  negligibly small for solutes below micro-metric size and  can be safely omitted.
One thus gets the PC formula
\begin{equation}
\Delta G_{\rm PC}^{\rm DFT} =  \Delta \Omega_{\rm DFT} - P_{\rm DFT} \Delta V_{\rm   DFT}.
\end{equation}
Note that the above equations rely on macroscopic thermodynamics and are strictly true for a macroscopic solute of volume $\Delta V$. They raises the question of the proper definition of the partial molar volume for a microscopic solute.
Numerical experiments for small molecular solutes have suggested that  the addition of an extra $\rho_0 kT \Delta V$ correction can  further improve the results in many cases. That is our PC+ correction \cite{sergiievskyi_fast_2014}. This  is also equivalent to reducing  the classical DFT pressure by an amount equal to the ideal pressure, $\rho_0 kT$:
\begin{equation}
\label{eq:DeltaG+}
 \Delta G_{\rm PC+}^{DFT} = \Delta 
 \Omega_{\rm DFT} - ( P_{\rm DFT} - \rho_0 kT) \Delta V_{\rm DFT}.
\end{equation}
We note that despite some arguments in Ref.\cite{sergiievskyi_fast_2014}, there is no clear justification for such factor and it even becomes contradictory for hydrophobic solutes of nanometer size (see next section). At this stage, it should be considered as an empirical adjustment of either the pressure, or of the solute partial molar volume at fixed pressure, for solutes of microscopic sizes. 

In the HRF (or HNC) approximation, which is commonly used in the DFT approach, the excess free energy functional $\mathcal{F}^{\rm exc}$ is expressed as a second-order Taylor series around the homogeneous fluid density $\rho_0$. It is important to note that    DFT and 3D-RISM share the same approximation in this case.
In a molecular-based framework (molecular DFT: MDFT),  the classical DFT functional  is written then as follows
:
\begin{eqnarray}
&& \Delta \Omega[\rho] = 
 \Omega[\rho] - \Omega[\rho_0]
 =
  kT
 \int
  \left[
  \rho(1) 
  \ln { \rho(1) \over \rho_0 }
  - \Delta \rho(1)
  \right]
  {\rm d1}
   \nonumber\\
   && +
  \displaystyle
  \int U(1) \rho(1) {\rm d1}
  -
  {kT \over 2}
  \iint  \Delta \rho(1) c(12) \Delta \rho(2) {\rm d1d2}\label{eq:DeltaOmega},
\end{eqnarray}
where the arguments $1$, $2$ stand for the positions and orientations of the solvent molecules,
$U$ is the external potential due to the solute molecule,
and $\Delta \rho(1) = \rho(1) - \rho_0$.
$c(12) = -\beta \delta^2 \mathcal{F}^{\rm exc} / \delta \rho(1) \delta \rho(2)$ is a pair direct correlation function of the pure solvent
at uniform density $\rho_0$ and at temperature $T$.
By minimizing the functional with respect to  the solvent density $\rho(1)$ one  finds both the solvation free energy $\Delta \Omega$ and the density distribution $\rho(1)$.

To apply the pressure correction we define the compressibility-route pressure of the theory, using the relation $\Omega[\rho_0] = -PV$. 
Insertion of the zero density $\rho=0$ into \eqref{eq:DeltaOmega} gives
\begin{equation}
\label{eq:PHNC}
P_{\rm DFT}  = \Delta \Omega[0]/V = \rho_0 kT -{kT \over 2} \rho_0^2  \hat c(k=0),
\end{equation}
where
$ 
 \hat c(k=0) = \int c(12) {\rm d1}
$, $\mathbf{k}$ is a Fourier-space coordinate.
Here and below we use the symbol ``$\hat{~~}$'' for the Fourier transformations of the real-space functions. 
The value of $\hat c(k=0)$ can be retrieved from all-atom simulations or from experiments by using, for instance, its relation to the isothermal compressibility $\kappa_T$ \cite{McQuarrie1976}: 
\begin{equation}
1 - \hat c(k=0) = \beta \kappa_T^{-1} \nonumber,
\end{equation}
with $\beta=\left(kT\right)^{-1}$. Finally, pressure corrections read:
\begin{eqnarray}
\label{eq:HNC/PC}
   &&\Delta G_{\rm PC}^{\rm DFT} =
   \Delta \Omega[\rho] 
    - \rho_0 kT( 1 -  {\rho_0 \over 2} \hat c(k=0)) \Delta V_{\rm DFT},\\
\label{eq:HNC/PC+}
   &&\Delta G_{\rm PC+}^{\rm DFT} =
   \Delta \Omega[\rho] 
    + {kT \over 2} \rho_0^2 \hat c(k=0) \Delta V_{\rm DFT}.
\end{eqnarray}

To test those formulae, we have plotted in Figure \ref{Classical DFT_vs_3D-RISM}  the  solvation free energies of 443 organic molecules in (SPCE) water using the classical DFT
functional for water and the classical DFT code developed by Jeanmairet, Levesque and Borgis \cite{jeanmairet_molecular_2013,jeanmairet_molecular_2013-1,jeanmairet_molecular_2015} (in the HNC approximation). Molecules and force fields are taken from \cite{david_l_mobley_small_2009}; the full list is given in supplementary information. We have also performed the same calculation using the 3D-RISM method with multi grid implementation of Sergiievskyi et al. \cite{sergiievskyi_3drism_2012,volodymyr_sergiievskyi_rism-mol-3d:_2013}.
In Table \ref{tab:Classical DFT_vs_3DRISM}, we give mean errors and standard deviations of both MDFT and 3DRISM with PC and PC+ corrections. PC+ halves the error of PC corrected SFE.

For  the pressure corrections,   we tried for both methods the two formulae \eqref{eq:HNC/PC} and \eqref{eq:HNC/PC+}. It can be seen that the results for classical DFT and 3D-RISM calculations differ. 
The best  DFT results are achieved using the formula \eqref{eq:HNC/PC+}, while the best 3D-RISM results correspond to the formula \eqref{eq:HNC/PC}.
This discrepancy is consistent with the findings of Misin et al. \cite{misin_communication:_2015} who advocated for the use of PC instead of PC$^+$ for a different data-base of molecules.

We anticipate at this point that the two series of results become consistent again if the pressure for bulk water in 3D-RISM is  defined as
\begin{equation}
\label{eq:P3D-RISM/water}
  P_{\rm 3DRISM} = 2 \rho_0 kT  -  \frac{kT}{2} \rho_0^2 \hat c(k=0)  
\end{equation}
so that the PC correction to 3D-RISM reads
\begin{equation}
\label{eq:RISM/PC}
   \Delta G_{\rm PC}^{\rm 3DRISM} = 
   \Delta \Omega_{\rm 3DRISM}[\rho] 
    - \rho_0 kT( 2 -  {\rho_0 \over 2} \hat c(k=0)) \Delta V_{\rm 3DRISM},
\end{equation}
and the PC+ correction reads
\begin{eqnarray}
\label{eq:RISM/PC+}
   \Delta G_{\rm PC+}^{\rm 3DRISM} &=& 
   \Delta \Omega_{\rm 3DRISM}[\rho] \nonumber\\
    &&- \rho_0 kT( 1 -  {\rho_0 \over 2} \hat c(k=0)) \Delta V_{\rm 3DRISM}.
\end{eqnarray}
In this case, PC+ for 3D-RISM would be equivalent to PC for MDFT. That would also explain the apparent difference between DFT and 3D-RISM in Figure \ref{Classical DFT_vs_3D-RISM} and in Ref \cite{misin_communication:_2015}.

It is the purpose of the next section to prove that the pressure expression given in Eq.~\ref{eq:P3D-RISM/water} is indeed the correct one for 3D-RISM.

\section{\label{sec:3D-RISM} Expression of the bulk solvent pressure in 3D-RISM}

3D-RISM equations for a one-component solvent with $n_s$ sites in (Fourier) $\mathbf k$-space can be written in the following form \cite{volodymyr_sergiievskyi_modelling_2013,hirata_molecular_2003}
\begin{equation}
\label{eq:3D-RISM}
 \hat{ \mathbf h}(\mathbf k) = \hat{\mathbf X}(|\mathbf k|) \hat{\mathbf c}(\mathbf k)
\end{equation}
where 
$\hat{ \mathbf h} = (\hat h_1(\mathbf k),\dots,\hat h_{n_s}(\mathbf k))^{\rm T}$,
$ \hat{\mathbf c} = (\hat c_1(\mathbf k),\dots,\hat c_{n_s}(\mathbf k))^{\rm T}$ are  the vectors of total and direct solute-solvent correlation functions. 
$\hat{\mathbf X}(k)$ is a matrix of susceptibility functions
\begin{equation}
 \hat{ \mathbf X}(k) =  \hat{\mathbf W}(k) +  \rho_0 \hat{\mathbf H}(k),
\end{equation}
where
$ \hat{\mathbf W}(k) = ( \hat \omega_{ij}(k) )$ is the matrix of intramolecular correlation functions, $\hat \omega_{ij}(k) = 
\sin(k r_{ij}) / k r_{ij}$, and 
$\hat{\mathbf H}(k) = (\hat h_{ij}^{solv}(k))$ is the  matrix of solvent-solvent correlation functions.

In the HNC approximation the 3D-RISM equations are completed by $n_s$ closure relations for $ i=1,\dots,n_s$ : 
\begin{equation}
\displaystyle
\label{eq:closure}
  g_i(\mathbf{r}) \equiv h_i(\mathbf{r}) + 1 = 
   \exp \left( -\beta u_i(\mathbf r) + 
         h_i(\mathbf r) - 
         c_i(\mathbf r)
         \right).
\end{equation}

\noindent From \eqref{eq:3D-RISM} we have $\mathbf c(\mathbf k) = \mathbf X^{-1}(k) \mathbf h (\mathbf k)$ and thus
\begin{equation}
  \mathbf h( \mathbf r) - 
  \mathbf c( \mathbf r) 
  =
  \rho_0
  \int \mathbf Z(|\mathbf r_2-\mathbf r|) \mathbf h(\mathbf r_2) d\mathbf r_2 ,
\end{equation}
where elements of the matrix 
$\mathbf Z(r)$ are the inverse 3D-Fourier transforms of the elements of the matrix $\hat{\mathbf Z}(k) \equiv  \rho_0^{-1}(\mathbf I_n - \hat{\mathbf X}^{-1}(k))$. 
Then from \eqref{eq:closure} we have
\begin{equation}
\label{eq:xclosure}
 \ln g_i(r) = -\beta u_i(\mathbf r) + \sum_j \rho_0 \int z_{ij}(|\mathbf r_2 - \mathbf r |) h_j(\mathbf r_2) d\mathbf r_2,
\end{equation}
where $z_{ij}(r)$ are the elements of $\mathbf Z(r)$.
It can be easily seen that expression \eqref{eq:xclosure} can be obtained by taking the functional derivative of the following 3D-RISM density functional $\mathcal F_{\rm 3DRISM}[\bm \rho_1,\dots,\bm \rho_n]$ over site-density $\rho_i(\mathbf r)$ 

\begin{eqnarray}
\label{eq:F3D-RISM}
&&\mathcal F_{\rm 3DRISM} [\bm \rho_1,\dots,\bm \rho_{n_s}] =\nonumber\\
&&\sum_i \left( kT \int  \rho_i(\mathbf r) 
      \ln \left(\frac{\rho_i(\mathbf r)}{\rho_0}\right)
      -
      \Delta \rho_i(\mathbf r)  {\rm d}\mathbf r
+ \int \rho_i(\mathbf r) u_i(\mathbf r) {\rm d} \mathbf r
\right)\nonumber\\
&&-\frac{kT}{2}\sum_{ij}
\iint \Delta \rho_i(\mathbf r_1) 
z_{ij}(|\mathbf r_2 - \mathbf r_1|) 
\Delta \rho_j(\mathbf r_2) 
{\rm d} \mathbf r_1 {\rm d} \mathbf r_2
\end{eqnarray}
and equating the derivative to zero. We use here the usual definitions: $\rho_i(\mathbf r)$ is the density of site $i$ at the position $r$, 
$\Delta \rho_i(\mathbf{r}) \equiv \rho_i(\mathbf{r}) - \rho_0$,
$g_i(\mathbf r) \equiv \rho_i(\mathbf{r}) / \rho_0$, and 
$h_i(\mathbf r) \equiv \Delta \rho_i(\mathbf r) / \rho_0$

\begin{figure}
\includegraphics[width=1\columnwidth]{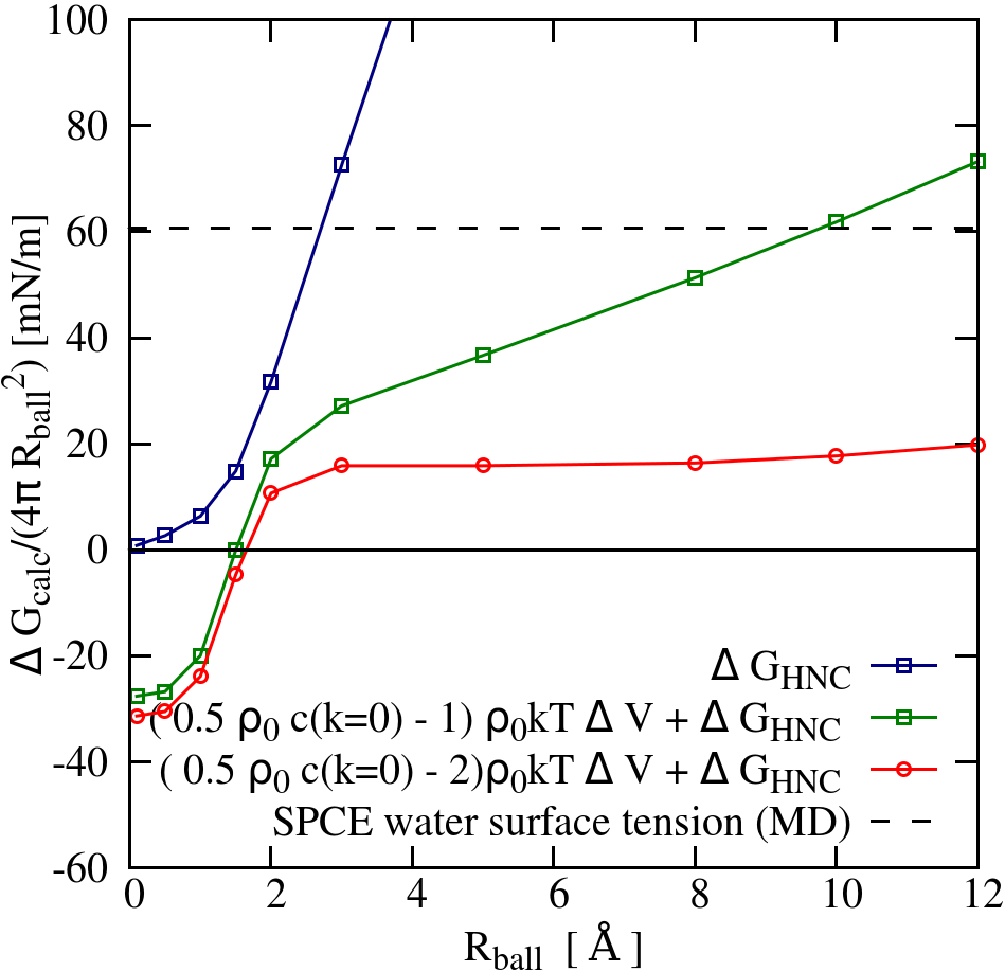}
\caption{
\label{fig:rism_surf_tens}
Solvation free energy per surface area for 3D-RISM. The correct behavior (a plateau for large radius) is achieved for the formula 
$\Delta G_{HNC} + \rho_0 kT ( {1 \over 2} \rho_0 \hat c(k=0) - 2) \Delta V$, the theoretical formula derived in Sec.~\ref{sec:3D-RISM}.
The MD-value of the surface tension of the non-flexible SPCE water (60.7 mN/m) is taken from Ref \cite{Yuet2010}.
}
\end{figure}

It should be noted, that this functional is similar, but not identical to the Site Density Functional introduced in Ref \cite{Liu_2013_siteDFT} and the difference is not only in the bridge, but also in the HNC part.
The functional in Ref \cite{Liu_2013_siteDFT} is constructed by  replacing of the six-dimensional direct correlation $c(12)$-function in \eqref{eq:DeltaOmega} by the sum of site-site functions, while the functional \eqref{eq:F3D-RISM} is derived to be consitent with the 3D-RISM equations \eqref{eq:3D-RISM}-\eqref{eq:closure}, which results in replacement of the sum of $c_{ij}$ by the sum of $z_{ij}$ functions, which are in turn dependent on the $(c_{ij})$ matrix.
Both approaches are valid and can be sucessfully used for the calculations. 
However, it is important to have in mind the difference in case of comparison of the Classical (site) DFT and 3D-RISM results.

This functional represents the grand potential difference between the system with and without the solvated molecule 
\begin{equation}
 \mathcal F_{\rm 3DRISM} [ \{\bm{\rho}_i\} ] 
 =
 \Omega_{\rm 3DRISM}
 [ \{\bm{\rho}_i\} ]
 -
 \Omega_{\rm 3DRISM}
 [ \rho_0 ],
\end{equation}
where $\{\bm{\rho}_i\} \equiv \bm \rho_1,\dots,\bm \rho_{n_s}$.
Using the thermodynamic relation for the bulk grand potential $\Omega = -PV$ and considering the case $\rho_1 = \dots \rho_{n_s} = 0$ we get
\begin{equation}
 P_{\rm 3DRISM}V = \mathcal F_{\rm 3DRISM}[0,\dots,0].
\end{equation}
Using this equation we find the expression of the bulk pressure in the 3D-RISM approximation
\begin{equation}
  P_{\rm 3DRISM} = n_s \rho_0 kT 
  - {kT \over 2} \rho_0^2 \sum_{ij} \hat z_{ij}(k=0),
\end{equation}
where $\hat z_{ij}(k=0) = \int z_{ij}(r) d\mathbf r$.
In the expression above the sum of $\hat z_{ij}$ functions at $k=0$ can be expressed through the molecular direct correlation function $\hat c(k=0)$ (see Appendix, equation \eqref{eq:z from c}). This gives the following expression for the compressibility-route pressure in 3D-RISM
\begin{equation}
\label{eq:P3D-RISM}
 P_{\rm 3DRISM} =  { n_s  + 1 \over 2} \rho_0 kT
  - {kT \over 2} \rho_0^2 \hat c(k=0).
\end{equation}
It is easily seen that for $n_s = 1$ this expression coincides with the classical DFT pressure expression \eqref{eq:PHNC}.
The equation for water with $n_s=3$ gives the pressure \eqref{eq:P3D-RISM/water}, and thus proves the expression of Eq.~\ref{eq:P3D-RISM/water} and the final considerations
of the previous section.

To check the validity of this  expression of the 3D-RISM pressure in the case of water, we have used a procedure  performed previously for classical DFT \cite{jeanmairet_molecular_2015,jeanmairet_molecular_2013} and consisting in measuring the solvation
free energy of a growing hard sphere (or bubble) in water, which should follow the following relation for large radii $R$
\begin{equation}
\gamma  = \lim_{R \rightarrow \infty} \left( \frac{\Delta \Omega}{4 \pi R^2}  - \frac{PR}{3} \right),
\end{equation}  
where $P$ is the bulk pressure and $\gamma$ is the liquid-gas surface tension. This enables to measure both $P$ and $\gamma$ for the model. In Figure \ref{fig:rism_surf_tens}, we show that the correct plateau behavior for the surface tension is observed for the 3D-RISM pressure of Eq.~\ref{eq:P3D-RISM/water}. For what concerns DFT, the pressure expression of Eq.~\ref{eq:PHNC} gives the incorrect behaviour, but this already known shortcoming of HNC has been addressed recently\cite{jeanmairet_molecular_2015}.

\section{Conclusion}

In this paper, we have derived the pressure correction for the classical DFT and 3D-RISM methods and shown they are different. For the classical DFT formulation, we propose a simpler formulation. For the case of 3D-RISM, we have here expressed
the 3D-RISM/HNC equations in a functional form and define a 3D-RISM density functional to be optimized.
Using that functional and basic thermodynamic relations, a compressibility-route expression of the pressure could be obtained.
This pressure in 3D-RISM differs indeed from the pressure in molecular DFT and depends on the number of sites of the solvent molecule. The theoretical expression  was also shown to be consistent with that obtained numerically by computing the solvation free-energy of a growing sphere (or bubble) and comparing with the expected behaviour for large, macroscopic spheres.
Using the pressure representation, a consistent expression for the pressure correction (PC) and modified pressure correction (PC+) were written for 3D-RISM.
It was shown that the modified correction approximately halves the error on SFE predictions of the original PC correction. This is consistent with the results of Ref. \cite{misin_communication:_2015}, but raises the question of a sound theoretical justification: Only the PC correction is fully justified in the limit of a macroscopic solutes (but not for solutes of molecular sizes).

It is thus now possible to apply a pressure correction to the 3D-RISM/HNC approximations with arbitrary multi-atomic solvent. We recommend to use these corrections in the computation of solvation free-energies with 3D-RISM.

\appendix
\section*{ Appendix   }

To express the $\sum \hat z_{ij}(k=0)$ using the $\hat c_{ij}$ functions we use the auxliarly vectors $\mathbf e$ which are the $n_s \times 1$ vectors comprised of ones:
\begin{equation}
  \mathbf e \equiv (\underbrace{1,\dots,1}_{n_s})^T.
\end{equation}
Using this definition the sum of elements of the  matrix $\mathbf Z$ is expressed as $\mathbf e^T \hat{\mathbf Z} \mathbf e$. The $n_s \times n_s$ matrix comprised by ones can be written as $\mathbf e \mathbf e^T$, and $\mathbf e^T \mathbf e = n_s$.
So, we write $\sum \hat z_{ij}(k=0)$ in a following form:
\begin{equation}
\label{eq:Z}
  \mathbf e^T \hat{\mathbf Z} \mathbf e =
  \mathbf e^T \rho_0^{-1}(\mathbf I_n - \hat{\mathbf X}^{-1}) \mathbf e =
   {n_s \over \rho_0} - {1 \over \rho_0} 
   \mathbf e^T \hat{\mathbf X}^{-1} \mathbf e.
\end{equation}

Form the RISM equations we have \cite{chandler_optimized_1972,volodymyr_sergiievskyi_modelling_2013,hirata_molecular_2003}:
\begin{equation}
 \hat{\mathbf H}(k) = \hat{\mathbf W}(k)  \hat{\mathbf C}(k) 
  \hat{\mathbf X}(k).
\end{equation}
Although at $k=0$ the matrices in the RISM equations are degenerate the equations can be inverted at any infinitesimal $k\to 0$:
\begin{equation}
\label{eq:C}
  \mathbf e^T \hat{\mathbf C}(k) \mathbf e = 
  \mathbf e^T 
  \hat{\mathbf W}^{-1}(k) \hat{\mathbf H}(k) \hat{\mathbf X}^{-1}(k) 
  \mathbf e.
\end{equation}
All site-site solvent total correlation functions $\hat h_{ij}^{solv}(k)$ tend to the molecular correlation function $\hat h(k=0)$.
This is clear if we look at equality $\hat h_{ij}^{solv}(k=0)=\int h_{ij}^{solv}(\mathbf r)d\mathbf r$ and remember that the total number of sites of each kind in the system should be the same as the total number of molecules. Thus we can use the approximation $\hat{\mathbf H}(k) \approx \hat h(k=0) \mathbf e \mathbf e^T$ which gives 
\begin{equation}
\mathbf e^T
\hat{\mathbf W}^{-1} \hat{\mathbf H}
\approx 
\hat h(k=0) \mathbf e^T \hat{\mathbf W}^{-1} \mathbf e \mathbf e^T.
\end{equation}
It can  be shown that the sum of the elements in $\hat{\mathbf W}^{-1}$ tends to 1 when $k \to 0$.
Indeed, the matrix $\hat{\mathbf W}(k) = (\sin (kr_{ij}) / kr_{ij} ) \to \mathbf e \mathbf e^T$.
It is easy to see that $\mathbf e$ is an eigenvector of $\mathbf e \mathbf e^T$ matrix with the eigenvalue $n_s$:
$\mathbf e \mathbf e^T \mathbf e = n_s \mathbf e $.
Although at $k=0$ the matrix $\hat{\mathbf W}$ is degenerate at any small $k \to 0$ it is invertible. The inverse matrix $\hat{\mathbf W}^{-1}(k)$ would have the eigenvalue $\lambda \to  1/n_s$ for the eigenvector $\mathbf x \to \mathbf e$:
\begin{eqnarray}
  \hat{\mathbf W}(k) \mathbf e \approx n_s \mathbf e  
  ~~\Rightarrow \\
  \mathbf e \approx n_s \hat{\mathbf W}^{-1}(k) \mathbf e
  ~~\Rightarrow  \\
 \mathbf e^T \mathbf e \approx 
  n_s \mathbf e^T \hat{\mathbf W}^{-1}(k) \mathbf e.
\end{eqnarray}
Because $\mathbf e^T \mathbf e = n_s$ we have $\mathbf e^T \hat{\mathbf W}^{-1} \mathbf e \approx 1 $ and 
$\mathbf e^T \hat{\mathbf W}^{-1}(k) \hat{\mathbf H}(k) \approx \mathbf e^T \hat h(k=0)$. Inserting  this into \eqref{eq:C} we have
\begin{equation}
  \mathbf e^T \hat{\mathbf C}(k) \mathbf e 
  \approx 
\hat  h(k=0) \mathbf e^T \hat{\mathbf X}^{-1}(k) \mathbf e.
\end{equation}
Using the RISM assumption we get 
$\mathbf e^T \hat{\mathbf C}(k) \mathbf e = \sum_{ij}\hat c_{ij}(k) = \hat c(k)$.
From the Ornstein Zernike equation  \cite{ornstein_acculental_1914}
\begin{equation}
  \hat h(k) = { \hat c(k) \over   1 -  \rho_0 \hat c(k)  }
\end{equation}
and thus
\begin{equation}
  \mathbf e^T \hat{\mathbf X}^{-1}(k) \mathbf e \to 
  { \hat c(k=0) \over \hat h(k=0) }  = 1 - \rho_0 \hat c(k=0).
\end{equation}
Then equation \eqref{eq:Z} gives
\begin{equation}
\label{eq:z from c}
\sum_{ij} \hat z_{ij}(k=0)  = 
 {n_s - 1 \over \rho_0} + \hat c(k=0).
\end{equation}

\end{document}